\documentclass[aps,pra,twocolumn,floatfix,floats,showpacs,superscriptaddress,raggedbottom]{revtex4-1}
\usepackage{graphicx,latexsym}
\usepackage{dcolumn}
\usepackage{amsmath}
\usepackage{amssymb,bm}
\usepackage{color}
\usepackage[normalem]{ulem}

\def\mb#1{\mbox{\boldmath$#1$}}

\def\sec#1{Sec.\ \ref{#1}}
\def\eq#1{Eq.\ (\ref{#1})}
\def\fig#1{Fig.\ \ref{#1}}

\usepackage{hyperref}
\hypersetup{
    pdfnewwindow=true,       
    colorlinks=true,         
    linkcolor=blue,          
    citecolor=blue,          
    filecolor=magenta,       
    urlcolor=black           
}

\begin{document}

\title{Competition of static magnetic and dynamic photon forces\\ in electronic transport through a quantum dot}

\author{Nzar Rauf Abdullah}
\email{nzar.r.abdullah@gmail.com}
\affiliation{Physics Department, Faculty of Science and Science Education, School of Science, University of Sulaimani, 
             Kurdistan Region, Iraq}
\affiliation{Science Institute, University of Iceland, 
             Dunhaga 3, IS-107 Reykjavik, Iceland}

\author{Chi-Shung Tang}
\affiliation{Department of Mechanical Engineering,
  National United University, 1, Lienda, Miaoli 36003, Taiwan}

\author{Andrei Manolescu}
\affiliation{Reykjavik University, School of Science and Engineering,
              Menntavegur 1, IS-101 Reykjavik, Iceland}

\author{Vidar Gudmundsson}
\email{vidar@hi.is}
 \affiliation{Science Institute, University of Iceland,
        Dunhaga 3, IS-107 Reykjavik, Iceland}

%

\begin{abstract}
We investigate theoretically the balance of the static magnetic and the 
dynamical photon forces in the electron transport through a quantum dot in 
a photon cavity with a single photon mode. 
The quantum dot system is connected to external leads and 
the total system is exposed to a static perpendicular magnetic field. 
We explore the transport characteristics through the system by tuning 
the ratio, $\hbar\omega_{\gamma} / \hbar\omega_c$, between the photon energy, $\hbar\omega_{\gamma}$, 
and the cyclotron energy, $\hbar\omega_c$. 
Enhancement in the electron transport with increasing electron-photon coupling 
is observed when $\hbar\omega_{\gamma} / \hbar\omega_c > 1$.
In this case the photon field dominates and stretches the electron charge distribution 
in the quantum dot, extending it towards the contacts area for the leads. 
Suppression in the electron transport 
is found when $\hbar\omega_{\gamma} / \hbar\omega_c < 1$, as the external magnetic field 
causes circular confinement of the charge density around the dot. 
\end{abstract}



\maketitle

%
%

\section{Introduction}
Electron systems coupled to a quantized electromagnetic field are common component 
of semiconductor and superconducting nanoscale devices.
An optoelectronic system is formed by adding electronic and the photonic sources.
In optoelectronic circuits, electrons inelastically tunnel between 
two connected systems~\cite{Kouwenhoven413.2000}.  
The characteristics of electron tunneling is modified by the electron-photon 
interaction influencing the electron motion~\cite{Ishibashi314.437,Shibata109.077401}. 
The electron tunneling is the so called photon-assisted transport (PAT)~\cite{Kouwenhoven50.2019}.
The PAT is inelastic electron tunneling in which the energy of electrons is changed by 
photon emission and absorption processes. These two processes can
enhance or suppress the electron transport~\cite{Stoof53.1050,Torres72.245339,PhysRevA.72.032330}.
A suitable electronic structure for investigation of PAT is a quantum dot because of it's potential application in 
quantum information processing~\cite{Imamog72.210,Loss.57.1998,DiVincenzo.309.2005,Nielsen.2010}.
The PAT of both charges~\cite{Ishibashi314.437} and spins~\cite{Souza.84.115322} in quantum dots 
showing enhanced transport has been investigated.

If the electronic system is exposed to an external perpendicular magnetic field, the electron motion is also 
influenced by the magnetic field. It may form edge states~\cite{ThomasIhn2010}, or 
localized states~\cite{PhysRevB.82.195325} leading to decreased conductivity. 
Magnetic field has been considered to control electron-switching processes in qubits~\cite{ApplPhysLett.79.14},
and to enable magnetic resonance imaging in biology~\cite{Wilde2005} to mention two totally different applications. 

In the presence of both external magnetic and photon fields, magneto-phototransport
emerges in which the electrons are influenced by both fields.
Magneto-phototransport has been studied in superconducting complementary split-ring resonators coupled to the cyclotron
transition of two-dimensional electron gases~\cite{PhysRevB.90.205309} where blue-shifting 
of polaritons due to the diamagnetic term of interaction Hamiltonian was observed.
In addition, the magneto-phototransport has been investigated in graphene coupled to 
cavity photons when the vacuum Rabi frequency is comparable to, or even larger than
the cyclotron transition of Dirac fermions~\cite{PhysRevLett.109.267403}. 
The magneto-phototransport has not been investigated in quantum dots in a 
photon cavity in the presence of several photons.

In a previous publication we investigated PAT in a quantum dot (QD) system coupled to
cavity photons~\cite{Nzar.25.465302}. In this work, 
we study magneto-phototransport in a QD system coupled to a photon cavity using a generalized master equation 
(GME) \cite{Vidar61.305}.
We assume a QD embedded in a two-dimensional quantum wire in an external perpendicular magnetic field.
The DQ system is weakly connected to external leads and strongly coupled to the photon cavity with a single photon mode.
We show how the external magnetic and photon fields influence the electron transport in the QD system. 
We consider the cavity initially containing two photons polarized either parallel ($x$-direction) or 
perpendicular ($y$-direction) to the direction of the electron transport.
For the $x$-polarization, the electron transport 
is enhanced when the cyclotron energy is smaller than the photon energy while a suppression in the transport is noticed 
in the case of the cyclotron energy larger than the photon energy. No such transport characteristic is found in the case of 
$y$-polarization due to the anisotropy of the central system and the energy chosen for the photon.

This paper is organized as following: In \sec{Sec:Model_and_Theory} description of the model and the theoretical 
formalism are shown. In \sec{Sec:Results_and_Conclusions} we present the results and conclusions.

\section{Model and Theory}\label{Sec:Model_and_Theory}

In this section, we introduce the Hamiltonian of the system 
and the potential that forms the quantum dot.
The QD system is exposed to a uniform perpendicular magnetic field
and is in a quantized electromagnetic cavity with a single photon mode. 
The electron-electron and the electron-photon interactions
are explicitly taken into account. The photons in the cavity are linearly polarized.
We use a non-Markovian generalized master equation to investigate 
the non-equilibrium electron transport in the system.

\begin{figure}[htbq]
  \includegraphics[width=0.4\textwidth]{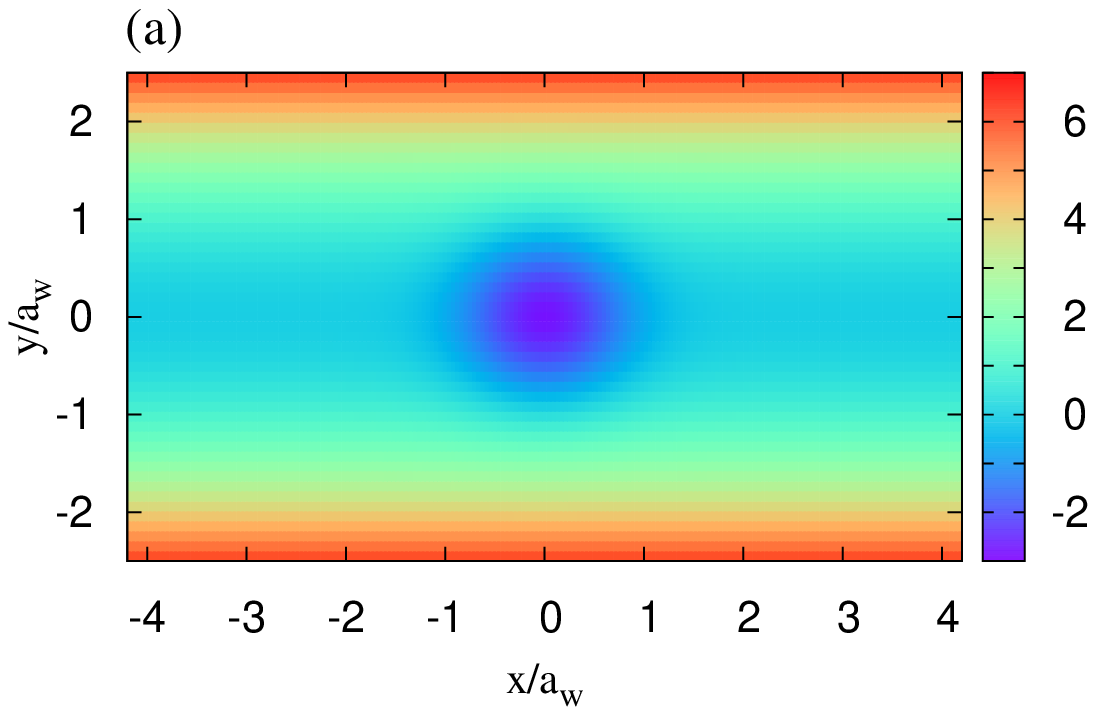}
  \includegraphics[width=0.4\textwidth]{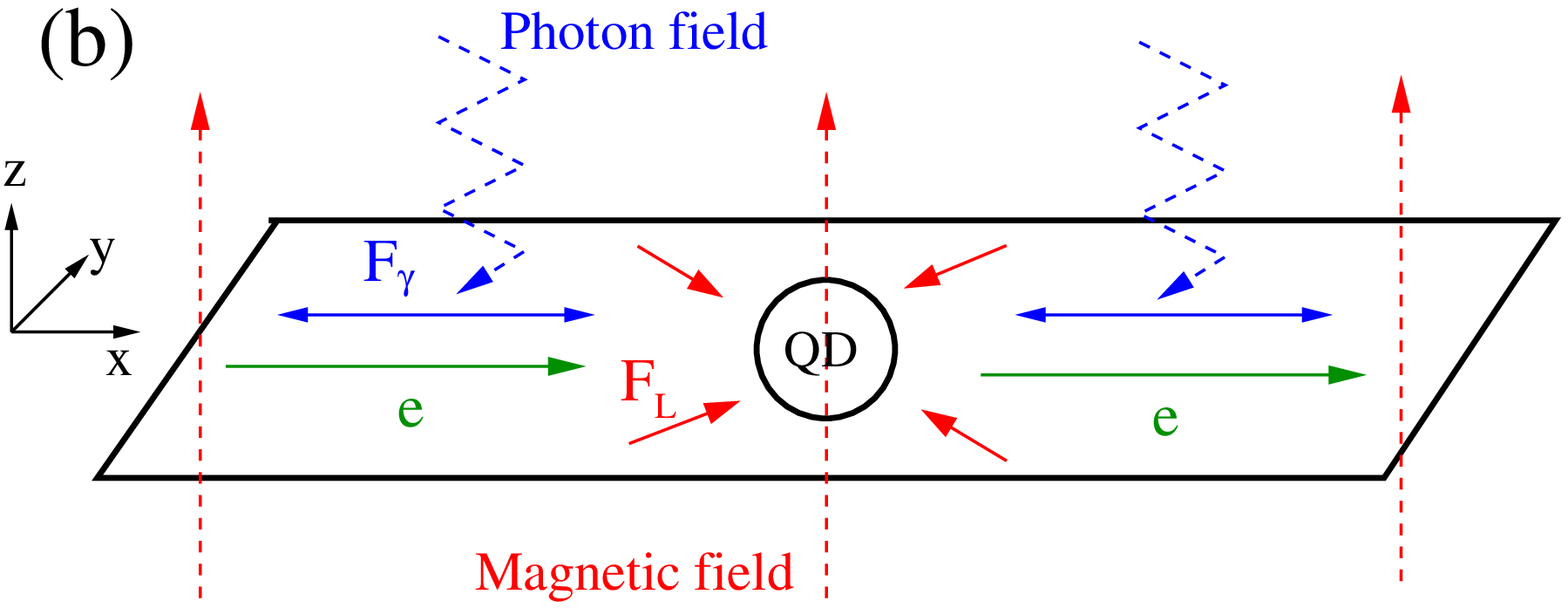}
 \caption{(Color online) (a) Potential of the quantum dot embedded in
    a quantum wire where $a_w$ is the effective magnetic length.
    The parameters $B = 0.1~{\rm T}$, $a_{w} = 23.8~{\rm nm}$, 
    and $\hbar \Omega_0 = 2.0~{\rm meV}$.
    (b) Schematic representation of the quantum dot system under influences of an external magnetic field 
    (red dashed line) and the photon field (blue dashed line). The magnetic field causes the Lorentz force 
    ($F_{\rm L}$) on the charge current and the photon field induces a force 
    ($F_{\gamma}$) on the charge density. The green arrows show the direction of
    electron (e) propagation in the system.}
\label{fig01}
\end{figure}

The central system is hard-wall confined in the $x$-direction and parabolocally confined in
the $y$-direction. The QD potential shown in \fig{fig01}(a) can be described by 
\begin{eqnarray}
 V_{\rm QD} = V_0 \exp(-\alpha_x^2 x^2 - \alpha_y^2 y^2),
\end{eqnarray}
where $V_0$ is the strength of the potential, and $\alpha_x$ and $\alpha_y$ are constant values
that define the diameter of the QD.

Figure \ref{fig01}(b) is a schematic representation of the QD system under the static external magnetic 
(red dashed line) and the dynamic photon field (blue dashed line). The magnetic field induces the Lorentz force 
($F_{\rm L}$), and the photon field induces a force ($F_{\gamma}$).

The Hamiltonian of the QD system coupled to a single photon mode 
in an external perpendicular magnetic field in the $z$-direction is \cite{2040-8986-17-1-015201,PhysRevE.86.046701}
\begin{align}\label{H_S}
 \hat{H}_{S}&=\int d^2r\;\hat{\psi}^{\dagger}(\mathbf{r}) \left[\frac{1}{2m^{*}}\left(\frac{\hbar}{i}\nabla+
              \frac{e}{c}\left[\mathbf{A}(\mathbf{r}) +\hat{\mathbf{A}}_{\gamma}(\mathbf{r})\right] \right)^2 \right. \nonumber \\
            & \left. + eV_{\rm pg} + V_{\rm QD}(\mathbf{r}) \right] \hat{\psi}(\mathbf{r}) + H_Z +
              \hat{H}_{ee} +\hbar \omega_{\gamma} \hat{a}^{\dagger}\hat{a}.
\end{align}
Herein, $\hat{\psi}$ is the field operator, $\mathbf{A}(\mathbf{r})= -By \hat{\mb{x}}$ is the vector potential 
of the external magnetic field defined in the Landau gauge, and
$\hat{\mathbf{A}}_{\gamma}$ is the vector potential of the photon field given by 
$\hat{\mathbf{A}}_{\gamma}(\mathbf{r})=A(\hat{a}+\hat{a}^{\dagger}) \mathbf{e}$ with
$A$ the amplitude of the photon field with the electron-photon coupling constant $g_{\gamma}=eA a_w \Omega_w/c$,
$\mathbf{e} = \mathbf{e}_x$ for parallel polarized photon field ($\mathrm{TE}_{011}$) or $\mathbf{e} = \mathbf{e}_y$ for 
perpendicular polarized photon field ($\mathrm{TE}_{101}$),
and $\hat{a}$($\hat{a}^{\dagger}$) annihilation(creation) operators of the photon in the cavity, respectively.
The effective confinement frequency is $\Omega_w = \sqrt{\Omega^2_0 + \omega^2_c}$ with $\Omega_0$ being electron confinement frequency
due to the lateral parabolic potential and $\omega_c$ the cyclotron frequency due to external magnetic field.
In addition, $V_{\rm pg}$ is the plunger gate voltage that controls the energy levels of the QD system
with respect to the chemical potential of the leads.
The second term is the Zeeman Hamiltonian describing the interaction between the external magnetic field and
the magnetic moment of an electron $H_Z = \pm g^{*}\mu_B B/2$ with $\mu_B$ the Bohr magneton and $g^{*} = -0.44$ the 
effective g-factor for GaAs.
The third term of \eq{H_S} ($\hat{H}_{ee}$)  
is the Coulomb repulsion between the electrons in the QD system \cite{Nzar.25.465302}.
Finally, the last term is the quantized photon field,
with $\hbar\omega_{\gamma}$ the photon energy.
%


The QD system is connected to two leads with the chemical potential of the left lead ($\mu_L$) 
higher than that of the right lead ($\mu_R$). The bias difference of the leads
causes electrons to be transferred from the left to the right lead when the system
has reached a steady state. 
The density operator of the $l$ lead before connection to the central system is
\begin{equation}
 \hat{\rho}_l = \frac{\exp(-\beta (\hat{H}_l-\mu_l \hat N_l))}{{\rm Tr}_l \{\exp{-\beta(\hat{H}_l-\mu_l \hat N_l)}\} },
\end{equation}
where $l \equiv \{L,R\}$, $\beta=1/(k_BT)$ with $k_B$ is the Boltzmann constant, $\hat N_l$ and $\hat{H}_l$ is the electron number operator
and the Hamiltonian of lead $l$, respectively~\cite{Vidar11.113007}.

The density operator of the total system
before coupling to the leads can be defined as the tensor product of the individual density operators
$\hat{\rho}(t<t_0) = \hat{\rho}_L \hat{\rho}_R \hat{\rho}_S(t<t_0)$.
Once the QD system and the leads are connected, the reduced density operator ($\hat{\rho}_{\rm S}$) 
describing the state of the electrons in the QD system under the influence of the leads
can be obtained by taking the trace over the Fock space of the leads
\begin{equation}
 \hat{\rho}_{\rm S}(t) = \rm{Tr}_{L+R}(\hat{\rho}).
\end{equation}

Now, we can calculate current carried by the electrons in the system from the reduced density operator.
We introduce net charge current ($I_{\rm Net}$)
\begin{equation}
 I_{\rm Net} = I_{L} + I_{R},
\end{equation}
where $I_{L}$ is the current from the left lead to the DQ system defined as
\begin{equation}
 I_{L}(t)=\mathrm{Tr}[\dot{\hat{\rho}}_S^L(t) \hat{Q}],
\end{equation}
where the charge operator is $\hat{Q} = e \hat{N}$ with the number operator $\hat{N}$.
The current from the QD system to the right lead ($I_{R}$) is 
\begin{equation}
 I_{R}(t)=-\mathrm{Tr}[\dot{\hat{\rho}}_S^R(t) \hat{Q}].
\end{equation}

In a steady state the right and left currents are of same magnitude. 
The time needed to reach the steady state depends on the chemical potentials in each
lead, the bias window, and their relation to the energy spectrum of the system.
In anticipation that the operation of an optoelectronic circuit can be sped up
by not waiting for the exact steady state we integrate the GME to $t=220$ ps, a point in
time late in the transient regim when the system is approaching the steady state.    

To show the dynamic motion of electrons in the central system 
the electron charge density is presented in the result section~\cite{Nzar.25.465302}.
Conclusions about the electron motion always needs the simultaneous checking of the
corresponding local current that will not be displyed here.  
%

\section{Results}\label{Sec:Results_and_Conclusions}
We study the effects of the external magnetic and photon fields on electron transport in a non-equilibrium system.
The QD system and the leads are made of GaAs
with electron effective mass $m^* = 0.067m_e$ and relative dielectric constant $\kappa = 12.4$.
The parameters that specify the radius of the dot are $\alpha_x = 0.03$~nm$^{-1}$, $\alpha_y = 0.03$~nm$^{-1}$, and $V_0 = -3.3$~meV.
The radius of the dot is thus $R_{\rm QD} \approx 33.33$~nm.
The cavity has a single photon mode with the photon energy $\hbar\omega_{\gamma} = 0.3$~meV,
and it initially contains two photons $N_{\gamma} = 2$.
The confinement energy of the QD system is equal to that of the leads 
$\hbar \Omega_0 = \hbar \Omega_l = 2.0$~meV. In addition, the temperature of the leads before coupling 
to the QD system is assumed to be $T = 0.001$~K (in order to avoid numerical instabilities at $T=0$).

\subsection{Photon cavity with $x$-polarization}\label{x-direction}

In this section we assume the photons in the cavity are polarized in the $x$-direction. 
We vary the cyclotron energy with the strength of the external magnetic field, 
$\hbar \omega_c = e\hbar B/m^*c$,
and fix the photon energy at $\hbar\omega_{\gamma} = 0.3$~meV.

\begin{figure}[htbq]
       \begin{center}
  \includegraphics[width=0.23\textwidth,angle=0,bb=54 60 215 294,clip]{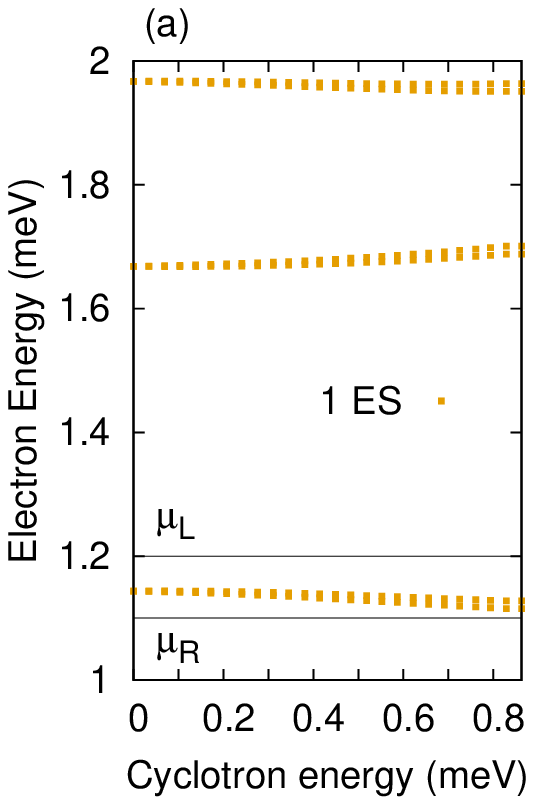}
  \includegraphics[width=0.23\textwidth,angle=0,bb=54 60 215 294,clip]{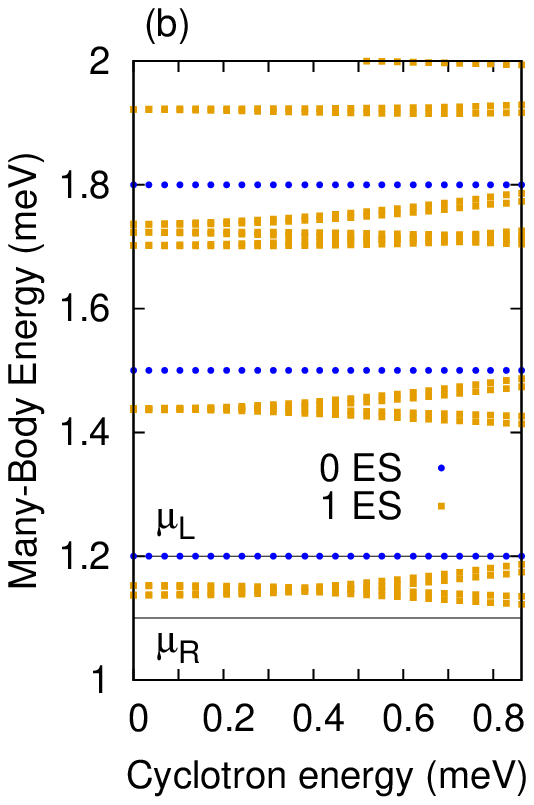}
         \end{center}
 \caption{(Color online) Energy spectra versus cyclotron energy plotted for the QD system 
           without (a) and with (b) the photon field, including zero-electron states (0ES, blue dots), 
           one-electron states (1ES, golden squares).
           The chemical potentials are $\mu_L = 1.2\ {\rm meV}$ and $\mu_R = 1.1\ {\rm meV}$ (black). 
           The plunger gate voltage $V_{\rm pg} = 0.4$~meV, $\hbar\omega_{\gamma} = 0.3$~meV, $g_{\gamma} = 0.10$~meV.
           The SE state in the bias window is almost doubly degenerate due to the small Zeeman energy.}
       \label{fig02}
\end{figure}

In order to calculate the energy spectrum of the QD-cavity system,
exact-diagonalization technique is utilized to diagonalize the matrix elements of 
the Hamiltonian \eq{H_S}.
Figure \ref{fig02} shows the energy spectra as a function of the cyclotron energy
for the QD system without (a) and with (b) the photon cavity, including zero-electron states 
(0ES, blue dots), one-electron states (1ES, golden squares). 
Two-electron states have higher energies because of the Coulomb repulsion.

The chemical potential of the left and the right leads (black lines) are
$\mu_L = 1.2$~meV and $\mu_R = 1.1$~meV, respectively. The bias window is thus $\Delta \mu = \mu_L - \mu_R = 0.1$~meV. 
The plunger-gate voltage is assumed to be $V_{\rm pg} = 0.4$~meV for the sake of putting the first-excited state into 
the bias window.

In \fig{fig02}(a) single-electron energy spectrum versus the cyclotron energy ($\hbar \omega_c$) is plotted.
The first-excited state is found in the bias window for the selected range of the cyclotron energy while the 
ground state is located below $1.0$~meV (not shown). At $V_{\rm pg} = 0.4$~meV, the first-excited state is 
getting into resonance with the first subband energy of the leads (not shown). 
Varying the cyclotron energy, the first-excited state stays in the bias window
and actively contributes to the electron transport in the QD system.

When the QD system is coupled to the photon cavity, extra photon replica states are formed for each ME state 
with different photon content as is displayed in \fig{fig02}(b). The energy difference between the photon replica 
states is approximately equal to the photon energy. For instance, the states around $1.40{\text-}1.45$~meV and 
$1.70{\text-}1.76$~meV are the photon replicas of the first-excited state containing approximately one and two 
photon(s), respectively. Other photon replicas of the ground-state around $1.18$~meV, $1.49$~meV and $1.78$~meV 
with photon content $\sim 4$, $5$, and $6$ are seen, respectively. 
But they do not participate to the transport because the cavity initially contains
two photons. In fact, the photon replica states play an essential role in the electron transport as we will 
be able to state later. 

\begin{figure}[htbq]
  \includegraphics[width=0.5\textwidth]{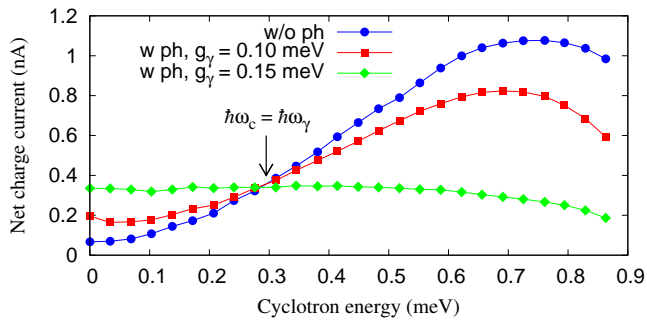}
 \caption{(Color online) The net charge current is plotted as a function of the 
cyclotron energy at time $t = 220$~ps for the QD system without photon (w/o ph) (blue circles), 
and with photon (w ph) cavity in the case of the electron-photon coupling strength $g_{\gamma} = 0.10$~meV (red squares)
and $0.15$~meV (green diamonds). The plunger gate voltage is $V_{\rm pg} = 0.4$~meV, $\hbar\omega_{\gamma} = 0.3$~meV, and
$\Delta \mu =0.1~{\rm meV}$.}
\label{fig03}
\end{figure}

Figure \ref{fig03} shows the net charge current versus the cyclotron energy ($\hbar\omega_c$)
at $t = 220$~ps and $V_{\rm pg} = 0.4$~meV for the QD system without a photon (w/o ph) (blue circles), 
and with a photon (w ph) cavity in the case of the electron-photon coupling $g_{\gamma} = 0.10$~meV (red squares) 
and $0.15$~meV (green diamonds). The photon energy is fixed at $\hbar\omega_{\gamma} = 0.3$~meV.

\begin{figure*}[htbq]
  \begin{center}
  \includegraphics[width=0.2\textwidth,angle=0,bb=220 59 410 300]{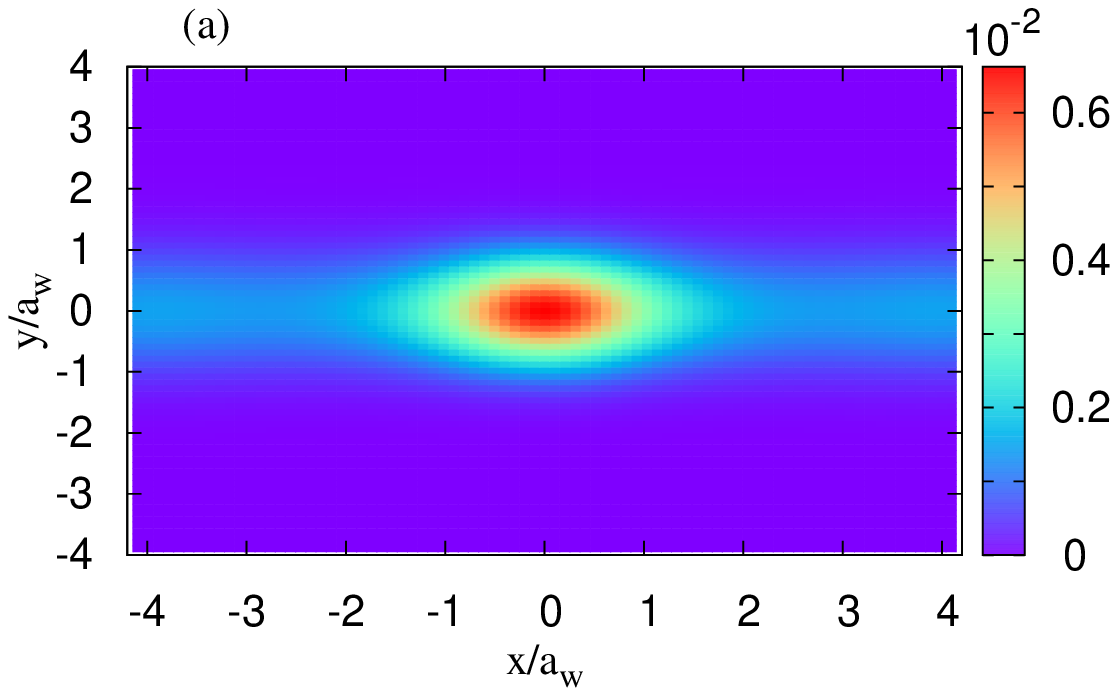}
  \includegraphics[width=0.2\textwidth,angle=0,bb=60 59 240 300]{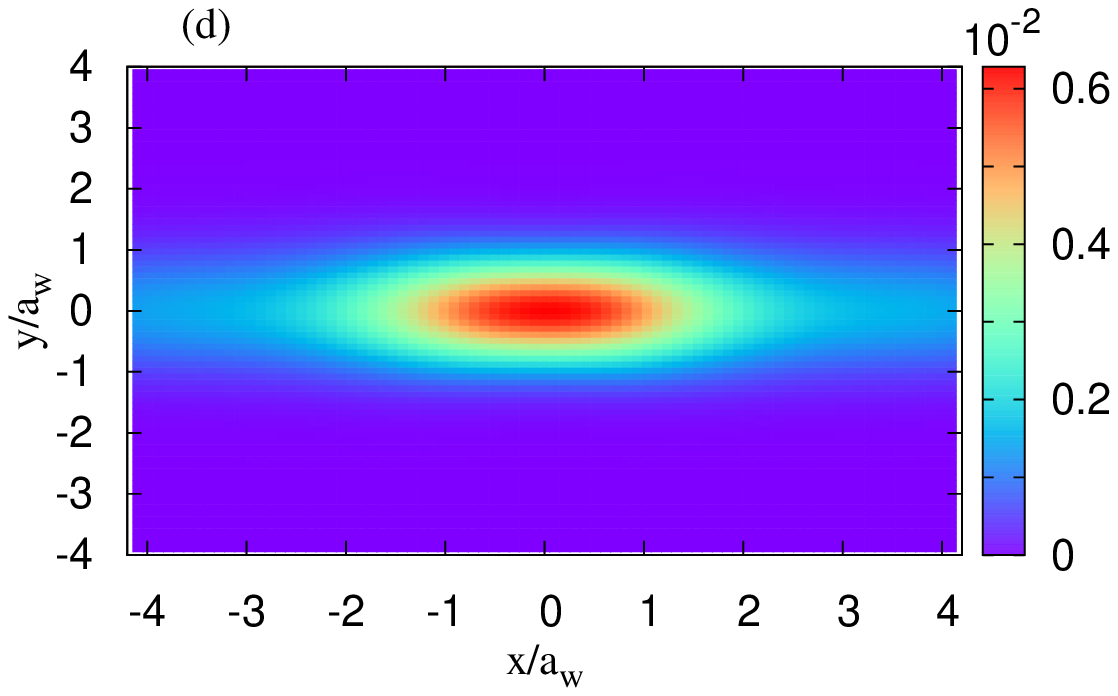}\\
  \includegraphics[width=0.2\textwidth,angle=0,bb=220 59 410 250]{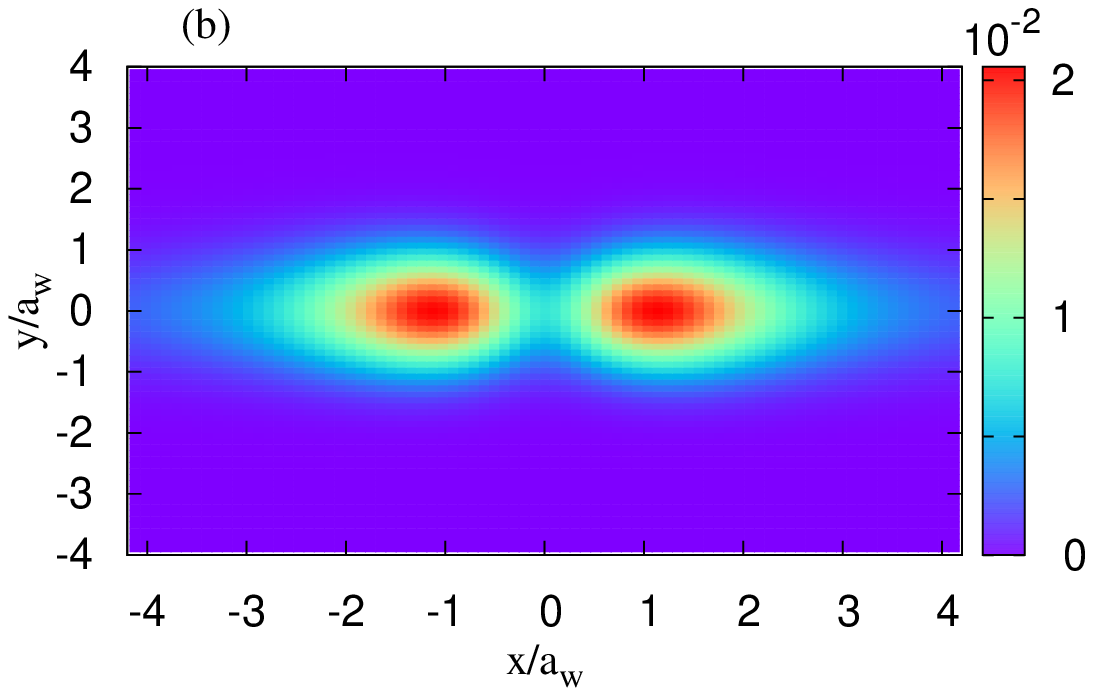}
  \includegraphics[width=0.2\textwidth,angle=0,bb=60 59 240 250]{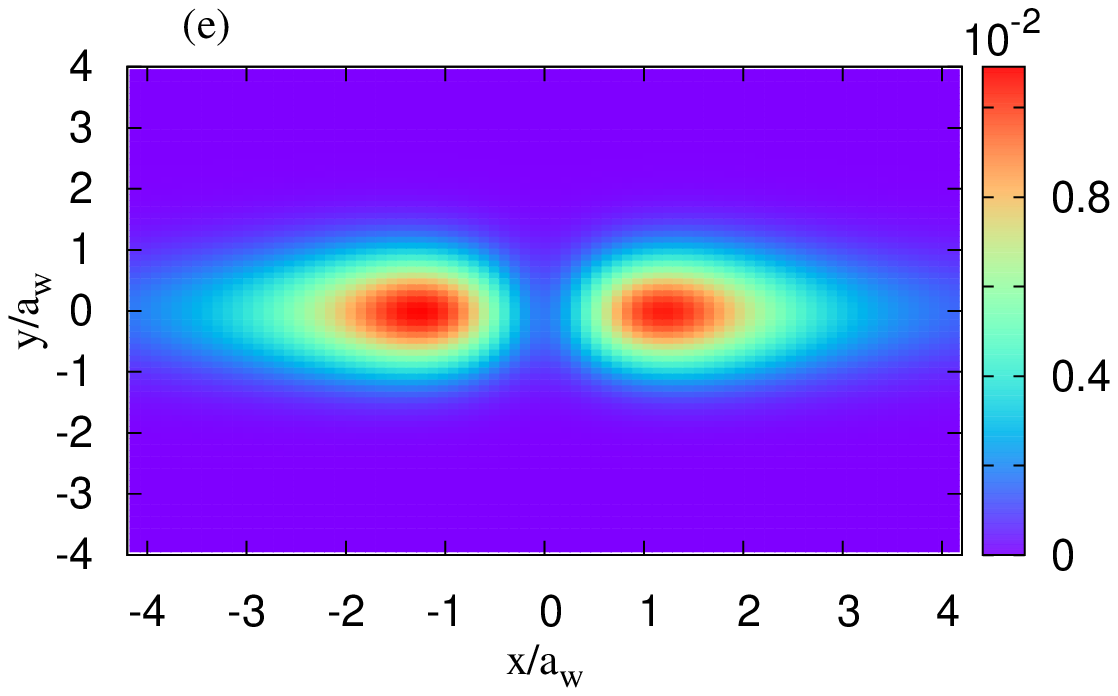}\\
  \includegraphics[width=0.2\textwidth,angle=0,bb=220 59 410 250]{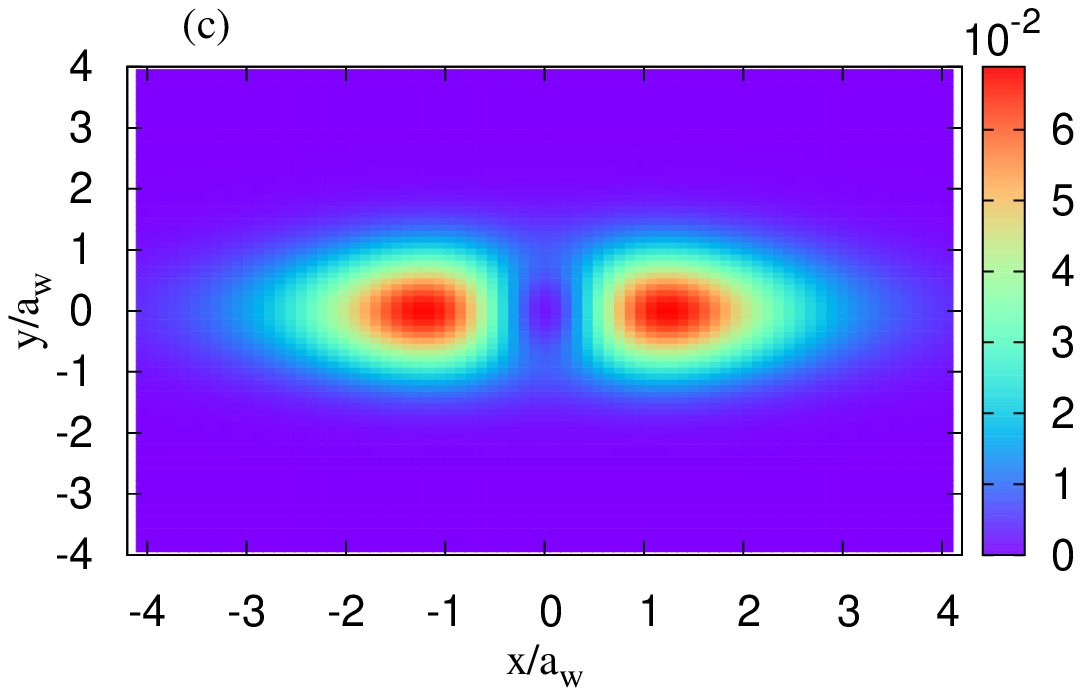}
  \includegraphics[width=0.2\textwidth,angle=0,bb=60 59 240 250]{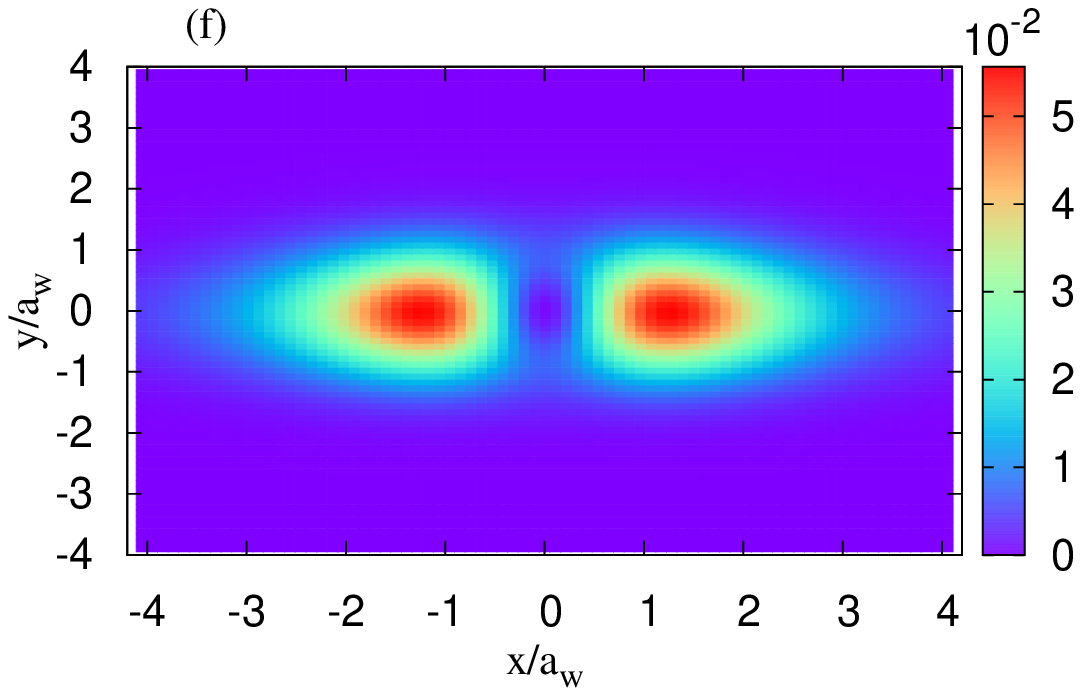}
   \end{center}
 \caption{(Color online) Distribution of charge density at $t = 220$~ps of the QD system without (left panels) and with (right panels)
         the photon field for three different cyclotron energy $\hbar\omega_c \simeq 10^{-4}$~meV (a)-(d), $0.3$~meV (b)-(e), 
         and $0.85$~meV (c)-(f). Other parameters are $\hbar\omega_{\gamma} = 0.3$~meV, $g_{\gamma} = 0.10$~meV,
         $L_x$ = $300$~nm, and $\Delta \mu =0.1~{\rm meV}$.}
\label{fig04}
\end{figure*}

In the absence of a photon field, 
the electrons from the first subband of the leads located in the bias window
may elastically tunnel to the first-excited state of the QD system
inducing current in the system. The net charge current is enhanced by increased cyclotron energy (blue circles). 
The electron charge density at the small cyclotron energy such as $\hbar \omega_c \simeq 10^{-4}$~meV
is localized in the QD and slightly distributed into the contact area to the leads as is seen in \fig{fig04}(a). 
The charge localization leads to charging of the QD system from both leads.
Consequently, the right current cancels out the left current and
the net charge current drops. By tuning the cyclotron energy to higher values such as $\hbar \omega_c \simeq 0.3$~meV and 
$0.85$~meV the charge density is further distributed outside the QD and a circular charge current emerges around the dot, 
due to the Lorentz force, as is shown in \fig{fig04}(c). 
As a result, the net charge current is enhanced compared to the case of lower cyclotron energy.
So, the electrons in the presence of the external static magnetic field are affected by the Lorentz force which 
shrinks the electron charge distribution towards the quantum dot, but does not localized it in the dot.

When we couple the QD system to the photon cavity, the first-excited state in the bias window, shown in \fig{fig03}(b),
is not active anymore, but instead photon replicas of it are active.
The influence of the external magnetic and the photon fields on the electron transport depends on their energies. 
Now, we seek the transport characteristics at three different energy ranges:
First $\hbar \omega_c < \hbar \omega_{\gamma}$, Second $\hbar \omega_c \simeq \hbar \omega_{\gamma}$, And 
third $\hbar \omega_c > \hbar \omega_{\gamma}$.

In the case of $\hbar \omega_c < \hbar \omega_{\gamma}$, 
the net charge current is enhanced as is demonstrated in \fig{fig03} for both electron-photon 
coupling strength $g_{\gamma} = 0.10$~meV (red squares) and $0.15$~meV (green diamonds).
The current enhancement is due to the following reasons: First, the activation of photon replica 
states in the transport. The photon replica of the first-excited state with approximately two photons around 
$1.70\text{-}1.76$~meV is the most active state and the weight of it's contribution to the electron transport 
is $70\%$. The second active state is the photon replica of the first-excited state containing one photon at 
$1.40\text{-}1.45$~meV which contributes $20\%$. The contribution of photon replica states enhances 
the net charge current because the higher states in the energy spectrum 
are less bound in the system.
The second reason is that the photon cavity induces extra `forces' on the charge density in the QD system, 
the `para- and the diamagnetic forces' \cite{2040-8986-17-1-015201}. 
The photon forces may stretch the electron charge in the central system 
towards the leads as displayed in \fig{fig04}(d).
Comparing to the charge density in the absence of the photon field shown in \fig{fig04}(a),
the charge density is affected by the photons, and therefore, the net charge current increases.

When the cyclotron energy is approximately equal to the photon energy  $\hbar \omega_c \simeq \hbar \omega_{\gamma}$
the effect of the Lorentz force matches that of the `para- and diamagnetic forces'
in the case of nonvanishing electron-photon coupling. 
The charge density is thus not effectively changed by the photon field as is displayed in \fig{fig04}(e).
The photon force includes the influences of the `para- and diamagnetic forces'.
Therefore, the net charge current in the QD system without a photon field around 
$\hbar\omega_c \simeq 0.3$~meV is almost equal to the net charge current in the presence of the photon cavity.

In the third case when the cyclotron energy is larger than the photon energy $\hbar \omega_c > \hbar \omega_{\gamma}$, 
the Lorentz force is dominant.
The contribution of the photon replica state with two photons is increased to $93\%$ while the photon replica 
state containing one photon is decreased to $3\%$. Blocking of the one-photon state in the transport is due to 
increased energy spacing between photon replicas of the states at a high cyclotron energy. 
The photon replica state with two photons is closer to getting into resonance  
increasing the net charge current comparing to that of the low cyclotron energy.
On the other hand, in the presence of the photon field a suppression in the current is seen at a high cyclotron 
energy compared to the net charge current in the absence of the photon cavity. The reason is that the increased Lorentz 
force induces a circular motion and collects the charge density around the QD as is shown in \fig{fig04}(f). 
As a result the net charge current is suppressed.

\begin{figure}[htbq]
       \begin{center}
  \includegraphics[width=0.25\textwidth,angle=0,bb=56 60 215 305,clip]{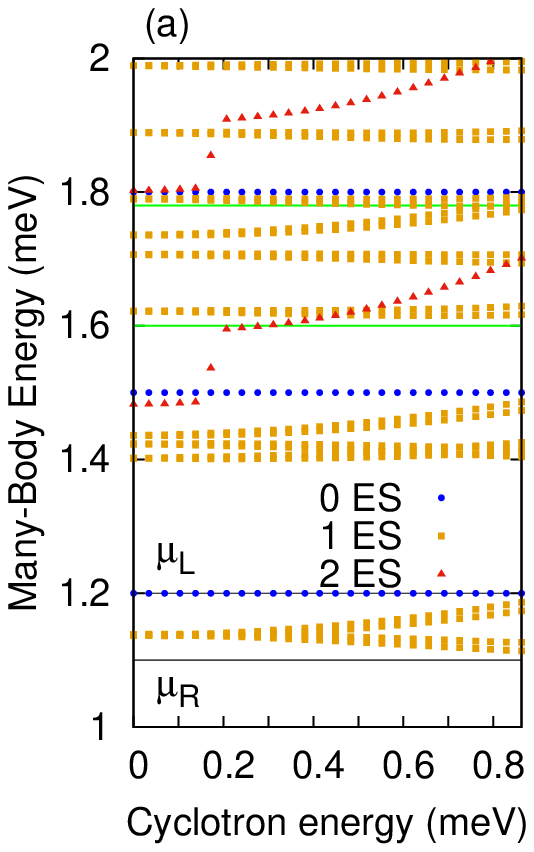}\\
  \includegraphics[width=0.45\textwidth]{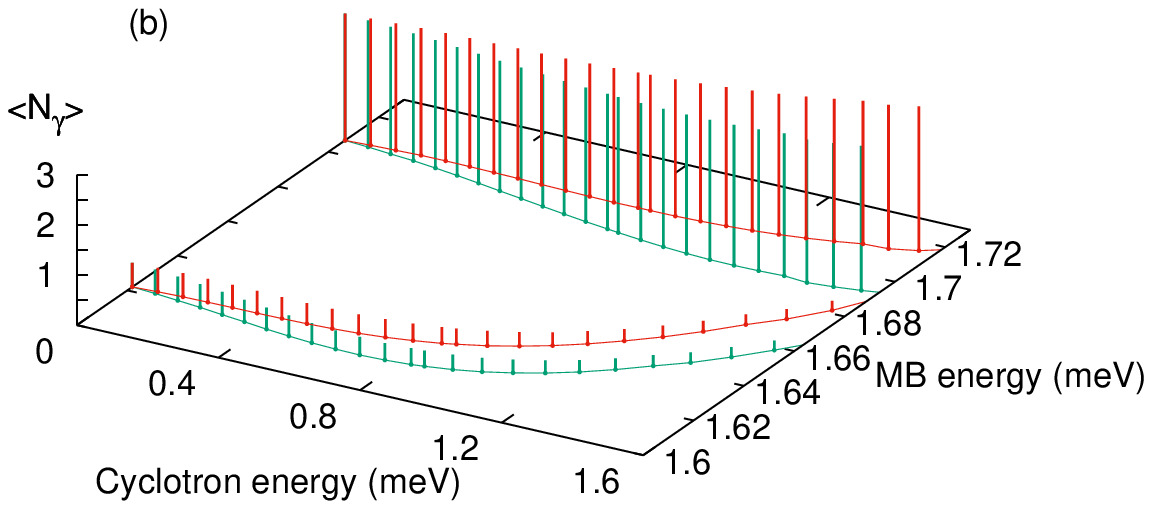}
         \end{center}
 \caption{(Color online) (a) MB Energy as a function of the cyclotron energy at $V_{\rm pg} = 0.1$~meV
           including zero-electron states (0ES, blue dots), one-electron states (1ES, golden squares)
           and two-electron states (2ES, red triangles) in the presence of the photon cavity.
           The chemical potentials are $\mu_L = 1.2\ {\rm meV}$, $\hbar\omega_{\gamma} = 0.3$~meV 
           and $\mu_R = 1.1\ {\rm meV}$ (black). 
           (b) The entanglement of both spin directions of the third-excited state $|31),|32)$ and the 
               photon replica of the first-excited state with approximately `three' photons $|33),|34)$.
               The photon content is shown with green bars (spin up) and red bars (spin down). The photon content
               is decreasing in $|31),|32)$ with increasing cyclotron energy, but it is increasing in the states $|33),|34)$.
               The SE state in the bias window is almost doubly degenerate due to the small Zeeman energy.}
       \label{fig05}
\end{figure}

To show further the importance of the photon replica states on the electron transport, 
we tune the plunger-gate voltage to $V_{\rm pg} = 0.1$~meV 
in order to shift down the energy spectrum. Figure \ref{fig05}(a) shows the MB energy spectrum versus 
the cyclotron energy at $V_{\rm pg} = 0.1$~meV for the QD system in a photon cavity.
The chemical potential of the left and the right leads (black lines) are
$\mu_L = 1.2$~meV and $\mu_R = 1.1$~meV, respectively.
It is clearly seen that the one-photon replica of the first-excited state  
enters into the bias window and the rest of the energy spectrum is shifted down by $0.3$~meV.
In this case, the current is totally due to states with more than one photon, because all the states within the bias window are 
photon replica states. The net charge current would be zero in the absence of a cavity.

We now expect the photon replica of the first-excited state with approximately three photons to be participating
in the transport. But three active one-electron states located between the green lines are found: 
The third-excited state in the energy range $\sim 1.60\text{-}1.63$~meV which is out of resonance. 
The photon replica of the first-excited state with photon content $2.53$ around $1.69\text{-}1.71$~meV, 
and the photon replica of second-excited state containing approximately one photon at $\sim 1.74\text{-}1.77$~meV.
In the low cyclotron energy range $\hbar \omega_c < \hbar \omega_{\gamma}$ all three channels mentioned above 
are active in the electron transport with a contribution of $30\%$ for each state. 
This is because the photon replica of first-excited state with approximately 
three photons is entangled with the third-excited state. Activation of a photon replica of the first-excited state 
causes participation of the third-excited state in the electron propagation.
The mean value of photons as a function of cyclotron energy and MB energy for the two entangled states is shown in \fig{fig05}(b).
We see that the photon content of the third excited state decreases by tuning the cyclotron energy to higher values
while the photon content of the photon replica of the first-excited state increases. 
We expect that whenever the contribution of one of these two states to the transport is getting weak,
the other state becomes weak as well.

\begin{figure}[htbq]
  \includegraphics[width=0.5\textwidth]{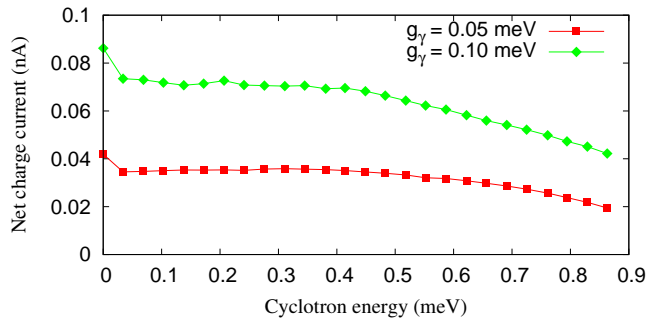}
 \caption{(Color online) The net charge current as a function of the
      cyclotron energy  plotted at time $t = 220$~ps and $V_{\rm pg} = 0.1$~meV in the presence of a photon cavity
      with the electron-photon coupling strength $g_{\gamma} = 0.05$~meV (red squares) and 
      $g_{\gamma} = 0.10$~meV (green diamonds) in the case of $x$-polarization of the photon field.
      The bias window is $\Delta \mu =0.1~{\rm meV}$, $\hbar \omega_{\gamma} = 0.3$~meV.}
\label{fig06}
\end{figure}

Figure \ref{fig06} displays the net charge current versus the cyclotron energy at $V_{\rm pg} = 0.1$~meV
where the electron-photon coupling strength is $g_{\gamma} = 0.05$~meV (blue circles) and $0.10$~meV (red squares) 
in the $x$-polarized of the photon field with photon energy $\hbar\omega_{\gamma} = 0.3$~meV.
We see that the net charge current is suppressed by increased cyclotron energy. As we have mentioned, 
three states at a low cyclotron energy participate in the electron transport leading to record high net charge current. 
By tuning the cyclotron energy to $\hbar \omega_c \sim 0.85$~meV where the cyclotron energy is larger than the photon energy
$\hbar \omega_c > \hbar \omega_{\gamma}$, the photon content of the third-excited state is approaching zero. 
So, the third-excited state is blocked.
The photon replica of the first-excited state with approximately three photons is also
getting weak in transport because these two states are entangled.
So, the main active state in the transport is the photon replica of second-excited state containing one photon. 
Therefore, the net charge current drops. The characteristics of the net charge current at $V_{\rm pg} = 0.1$~meV 
are totally opposite to those of the net charge current at $V_{\rm pg} = 0.4$~meV where the 
current is enhanced by increased cyclotron energy.

\begin{figure}[htbq]
  \begin{center}
  \includegraphics[width=0.2\textwidth,angle=0,bb=130 59 320 300]{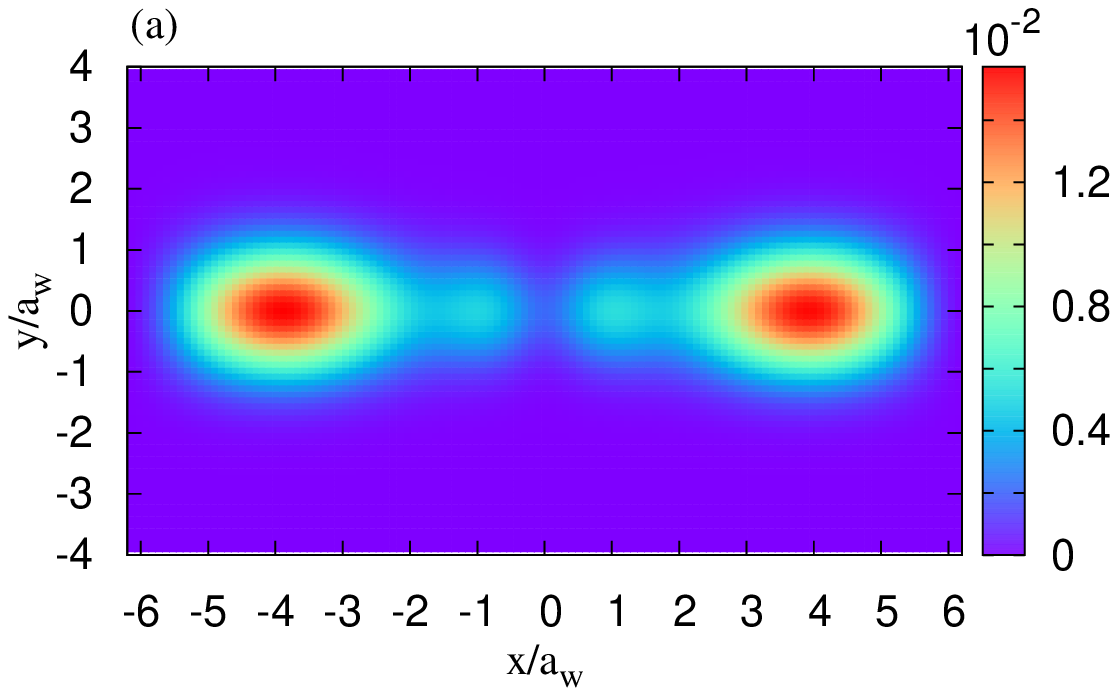}\\
  \includegraphics[width=0.2\textwidth,angle=0,bb=130 59 320 250]{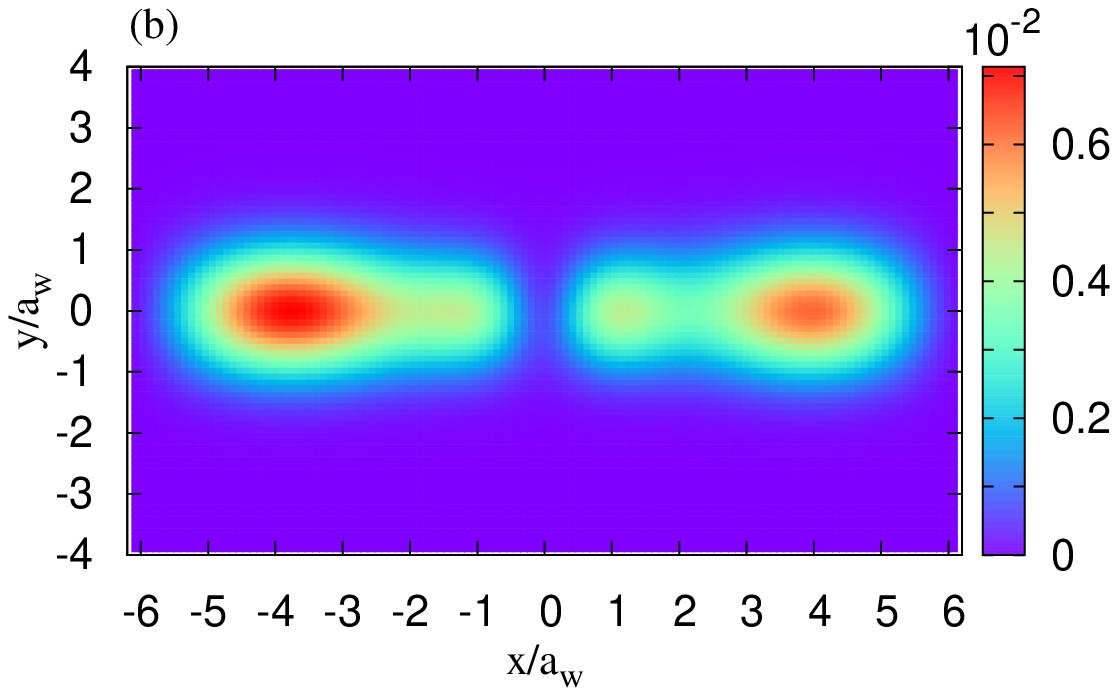}\\
  \includegraphics[width=0.2\textwidth,angle=0,bb=130 59 320 250]{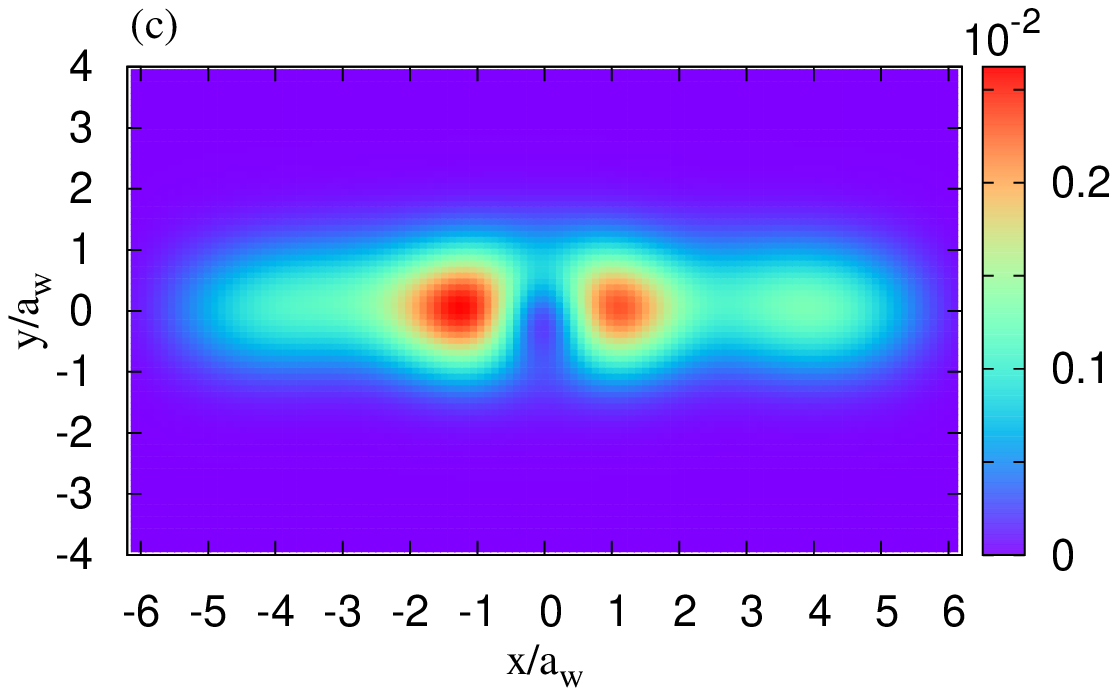}
   \end{center}
 \caption{(Color online) The charge density at $t = 220$~ps for the photon replica of 
 the first-excited state side-peak at $V_{\rm pg} = 0.1$~meV
 shown in \fig{fig06} for three different cyclotron energies $\hbar\omega_c \simeq 10^{-4}$~meV (a), 
 $0.34$~meV (b), and  $0.85$~meV (c) 
 in the presence of $x$-polarized photon field. Other parameters are 
 $\hbar\omega_{\gamma} = 0.3$~meV, $g_{\gamma} = 0.10$~meV,
 $N_{\gamma} = 2$, $L_x$ = $300$~nm, and $\Delta \mu =0.1~{\rm meV}$.}
\label{fig07}
\end{figure}

Figure \ref{fig07} displayes the charge density at $\hbar\omega_c \simeq 10^{-4}$~meV (a), $0.3$~meV (b), and  $0.85$~meV (c)
for the plunger gate $V_{\rm pg} = 0.1$~meV and time $t = 220$~ps.
In \fig{fig07}(a) the charge density is mostly distributed in the contact area to the leads indicating the 
dominance of the `para- and diamagnetic forces' which cause it's stretching.

By increasing the cyclotron energy to $\hbar\omega_c \simeq 0.85$~meV the Lorentz force is stronger than the 
`para- and diamagnetic forces'. The charge density indicates a circular motion around the QD 
as is demonstrated in \fig{fig07}(c). Consequently, the net charge current is reduced.

In the following, we shall show the influence of the photon cavity in the $y$-polarized photon field.

\subsection{Photon cavity with $y$-polarization}

\begin{figure}[htbq]
  \includegraphics[width=0.5\textwidth]{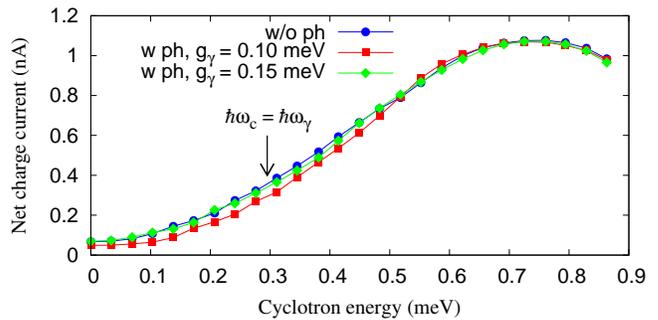}
 \caption{(Color online) The net charge current as a function of the 
cyclotron energy is plotted at time $t = 220$~ps and plunger gate voltage $V_{\rm pg} = 0.4$~meV
for the QD system without photon (w/o ph) (blue circles) and with photon (w ph) cavity with $y$-polarization 
in the case of $g_{\gamma} = 0.10$~meV (red squares) and $g_{\gamma} = 0.15$~meV (green diamonds). 
The bias window is $\Delta \mu =0.1~{\rm meV}$, $\hbar \omega_{\gamma} = 0.3$~meV, $N_{\rm \gamma} = 2$}
\label{fig08}
\end{figure}

In this section, we assume the photons in the cavity are polarized in the $y$-direction and the photon energy is fixed at 
$\hbar\omega_{\gamma} = 0.3$~meV with $\langle N_{\gamma}\rangle \approx 2$. The energy energy spectrum of the central system 
in the $y$-polarized photon field
is very similar to that of the $x$-polarization displayed in \fig{fig02}(b).
Figure \ref{fig08} shows the net charge current versus the cyclotron energy at $t = 220$~ps and 
$V_{\rm pg} = 0.4$~meV of the QD system without (w/o) (blue circles) and with photon (w ph) cavity in the case of
$g_{\gamma} = 0.10$~meV (red squares) and $g_{\gamma} = 0.15$~meV (green diamonds). 
The net charge current in the absence of the cavity
was explained in section \ref{x-direction}. 
Now we can clearly see that the net charge current is not influenced by the $y$-polarized photon field
for this selected photon energy due to the anisotropy of the system. The photon energy is far from resonance
with any electron state representing excitations in the $y$-direction.  

\section{Conclusions}

We have investigated the influences of a static magnetic and a dynamic photon fields on transport of electron through
a quantum dot system in a quantized photon cavity in an external perpendicular magnetic field. 
The photons are assumed to be polarized either parallel or perpendicular to the electron propagation in the QD system. 
The quantum dot system is connected to two leads and a non-Markovian master equation is used to describe 
time evolution of the electrons in the system~\cite{Vidar11.113007}. The motion of electrons in the 
system is influenced by the Lorentz force 
caused by the external magnetic field
and the `para- and diamagnetic forces produced by the photon field.  We have studied the characteristics 
of the net charge current in the system for three different cases: 
First, the `para- and diamagnetic forces' are dominant where the cyclotron energy is less than the photon energy, 
Second, the static and the dynamic forces are approximately `equal' when the cyclotron energy is 
`equal' to the photon energy, 
Third, the Lorentz force is dominant where the cyclotron energy is larger than the photon energy. 
In the first situation, the net charge current is enhanced because the dia- and paramagnetic forces extend 
the electron charge outside the quantum dot into the contact are of the external leads. 
In the third case an opposite situation happens because the Lorentz 
force tends to shrink the electron charge towards the quantum dot inducing a circular motion. 
Consequently, the net charge current is suppressed.

This investigation shows that the transport through a cavity photon system depends on
the balancing of the external forces and not only on the question which energy levels are 
close to resonance with the photon energy. Essential in these phenomena is the response of the 
charge density, or the information in the wavefunctions, to 
the external forces and thus the geometry of the system.  

\begin{acknowledgments}

Financial support is acknowledged from the Icelandic Research and Instruments Funds, 
and the Research Fund of the University of Iceland. The calculations were carried out on 
the Nordic high Performance computer (Gardar). We acknowledge the Nordic network
NANOCONTROL, project No.: P-13053, and the Ministry of Science and Technology, Taiwan 
through Contract No. MOST 103-2112-M-239-001-MY3.

\end{acknowledgments}

%

%

%
%
\end{document}